\pdfoutput=1
\documentclass[aps,prl,superscriptaddress,reprint,twocolumn]{revtex4-2}

\usepackage{amsmath,amssymb}
\usepackage{hyperref}
\hypersetup{pdfborder={0 0 0},colorlinks=true,urlcolor=blue,citecolor=blue,linkcolor=blue,linktocpage=true}

\newcommand{\nn}{\nonumber}
\newcommand{\dK}{\mathsf{K}}
\newcommand{\sugrarho}{\varrho}
\newcommand{\twistrho}{r}
\newcommand{\sugrasigma}{\sigma}
\newcommand{\twistsigma}{s}
\newcommand{\Vpot}{V_{\rm pot}}
\newcommand{\T}{\mathsf{T}}
\newcommand{\V}{\mathsf{v}}
\newcommand{\M}{\mathsf{m}}
\newcommand{\LL}{\mathsf{L}}
\newcommand{\coup}{\mathsf{g}}

\begin{document}

\title{Consistent Kaluza--Klein truncations and two-dimensional gauged supergravity}

\author{Guillaume Bossard}
\affiliation{Centre de Physique Th\'eorique, CNRS, Institut Polytechnique de Paris, 91128 Palaiseau cedex, France}

\author{Franz Ciceri}
\affiliation{Max Planck Institute for Gravitational Physics (Albert Einstein Institute), Am M\"uhlenberg 1, 14476 Potsdam, Germany}

\author{Gianluca Inverso}
\affiliation{INFN, Sezione di Padova, Via Marzolo 8, 35131 Padova, Italy}

\author{Axel Kleinschmidt}
\affiliation{Max Planck Institute for Gravitational Physics (Albert Einstein Institute), Am M\"uhlenberg 1, 14476 Potsdam, Germany}
\affiliation{International Solvay Institutes, ULB-Campus Plaine CP231, 1050 Brussels, Belgium}

\begin{abstract}
We consider generalized Scherk--Schwarz reductions of E$_9$ exceptional field theory to $D=2$ space-time dimensions and in particular construct the resulting scalar potential of all gauged supergravities that can be obtained in this way. This provides the first general expression for a multitude of theories with an interesting structure of vacua, covering potentially many new AdS${}_2$ cases. As an application, we prove the consistency of the truncation of eleven-dimensional supergravity on $S^8\times S^1$ to SO(9) gauged maximal supergravity. Fluctuations around its supersymmetric SO(9)-invariant vacuum  describe holographically the dynamics of interacting D0-branes.  
\end{abstract}

\maketitle

Flux compactifications of string theory are central in the AdS/CFT correspondence~\cite{Maldacena:1997re,Gubser:1998bc,Witten:1998qj} and in probing the quantum gravity swampland conjectures~\cite{Ooguri:2016pdq,Palti:2019pca}. Such compactifications lead to gauged supergravity theories in which some of the fields are charged under the vector fields.
The cases in which the classical solutions of gauged supergravity uplift consistently to solutions of ten- or eleven-dimensional supergravity are of particular interest.
The analysis of such solutions has allowed for many precision tests of the AdS/CFT correspondence in a large variety of AdS vacua, especially when the gauged supergravity has maximal supersymmetry, see e.g. \cite{Beisert:2010jr,Zarembo:2016bbk}. 
For instance, in the prototypical examples of the AdS/CFT correspondence the gravity side truncates consistently to an SO($N$) gauged maximal supergravity. 

The structure and dynamics of gauged maximal supergravities is well-understood in dimensions $D\ge3$ (see \cite{Samtleben:2008pe,Trigiante:2016mnt} for reviews). Their appearance as consistent truncations of a higher-dimensional parent theory is most efficiently analysed using the recent frameworks of generalized geometry~\cite{Lee:2014mla,Cassani:2019vcl} or exceptional field theory~\cite{Hohm:2014qga,Inverso:2017lrz,Galli:2022idq}.
The latter moreover allows for the derivation of the Kaluza--Klein spectrum and the analysis of the stability of the compactification~\cite{Malek:2019eaz,Guarino:2020flh}.

Gauged maximal supergravity in $D=2$ dimensions, by contrast, is less developed and only partial results are available~\cite{Nicolai:2000zt,Samtleben:2007an,Ortiz:2012ib}. 
At the same time, such theories are of particular interest in that most of their (supersymmetric) vacua are expected to contain an AdS${}_2$ factor with a running dilaton, a feature that has attracted attention recently in the context of applying the AdS/CFT correspondence to low-dimensional (Jackiw--Teitelboim)  gravity~\cite{Almheiri:2014cka,Maldacena:2016hyu,Maldacena:2016upp}.
A major example of the importance of $D=2$ is the conjectured holographic correspondence between solutions of SO(9) maximal gauged supergravity and the matrix model capturing the physics of the supermembrane modeled by stacks of D$0$-branes \cite{deWit:1988wri,Banks:1996vh,Sekino:1999av,Kanitscheider:2008kd,Ortiz:2012ib,Ortiz:2014aja}.

In this letter we report for the first time complete results for consistent truncations to $D=2$ gauged maximal supergravities.
In particular, we give the general expression for the scalar potential of these theories.
As an application we constructively prove the consistency of the truncation of the bosonic sector of type IIA supergravity on $S^8$ to SO(9) gauged maximal supergravity \cite{Ortiz:2012ib}, thus extending the partial uplift of the U(1)$^4$ invariant  
sector derived in \cite{Anabalon:2013zka}.
The full uplift of any solution of the SO(9) model back to ten or eleven dimensions can be derived from our expressions.

The original construction of gauged supergravities relied on a careful analysis of the supersymmetry transformations~\cite{deWit:1982bul,Hull:1984vg,Gunaydin:1985cu}, which can conveniently be phrased in the \textit{embedding tensor formalism}~\cite{Nicolai:2000sc,deWit:2002vt}.
This approach allows one to treat all possible gaugings on equal footing and to deal with expressions formally covariant under the global symmetry group of the original ungauged theory.
The process of turning part of the global symmetry into a gauge symmetry typically induces non-abelian interactions for the gauge fields. This deforms the Lagrangian and supersymmetry transformations and in particular introduces an intricate potential for the scalar fields at second order in the gauge coupling. The case of $D=2$ space-time dimensions has thus far resisted a comprehensive treatment from the point of view of supersymmetry due to the intricacies of the relevant representation theory~\cite{Nicolai:2004nv,Kleinschmidt:2021agj}.

In order to bypass the technical difficulties encountered in the supersymmetry analysis, in this letter we derive the scalar potential of $D=2$ gauged maximal supergravity by performing a generalized Scherk--Schwarz  reduction of the recently formulated E$_9$ exceptional field theory~\cite{Bossard:2018utw,Bossard:2021jix}.
Exceptional field theories capture the complete dynamics of ten- and eleven-dimensional supergravities in a form that is covariant under the E$_{n}$ groups that appear as global symmetries after a torus reduction to $D=11-n$ dimensions \cite{Julia:1980gr,Cremmer:1997ct}.
In particular, the  infinite-dimensional affine Kac--Moody extension E$_9$ of E$_8$ appears in $D=2$ dimensions~\cite{Julia:1982gx,Nicolai:1987kz}, where it acts on an infinity of scalar fields that are related by on-shell dualities.
Exceptional field theories are especially suited for studying  consistent truncations to gauged maximal supergravities through the aforementioned generalized Scherk--Schwarz reduction. The truncation ansatz is then mainly encoded in an E$_{n}$-valued `twist matrix' that determines the embedding tensor and  is subject to certain differential  constraints.
By construction, the potential of gauged supergravity only depends on the embedding tensor.
Exceptional field theory therefore provides an alternative route to identifying the $D=2$ scalar potential for any (upliftable) gauging without resorting to supersymmetry.

For brevity and in order to reduce technicalities, we restrict ourselves in this letter to the internal sector of the minimal formulation of E$_9$ exceptional field theory as defined in \cite{Bossard:2021jix}.
The results presented here can be generalized to include the full dynamics  of the extended formulation of the theory.

\subsection*{\texorpdfstring{Elements of E$_9$ exceptional field theory}{Elements of E9 exceptional field theory}}

E$_9$ exceptional field theory and geometry are  based on the loop algebra extension of the split real $\mathfrak{e}_8$, together with a Virasoro algebra acting on it~\cite{Bossard:2017aae,Bossard:2018utw,Bossard:2021jix}. Denoting the generators of E$_8$ by $T_0^A$ with $A=1,\ldots,248$, the loop extension allows for an arbitrary mode number $T^A_n$ with $n\in \mathbb{Z}$. We shall consider also the usual central extension by an element $\dK$ as well as the Virasoro generators $L_n$ with the standard relations. 
E$_9$ is generated by $\{T^A_n, \dK, L_0\}$. Following~\cite{Bossard:2018utw,Bossard:2021jix} we denote all these generators, including all $L_n$, collectively by $T^\alpha$ and define a set of (degenerate) bilinear forms $\eta_{k\, \alpha\beta}$ for $k\in\mathbb{Z}$ that pairs the loop generators $T^A_n$ and $T^A_{k-n}$ as well as $\dK$ and $L_k$~\footnote{For $k=0$ the form is related to the standard invariant bilinear form when restricted to the actual Kac--Moody algebra $\mathfrak{e}_9$.}.
Fields in E$_9$ exceptional field theory formally depend on infinitely many coordinates $Y^M$, taken from the so-called basic representation of E$_9$. This corresponds to the states of eight chiral bosons moving freely on the torus that is obtained by identifying points according to the E$_8$ root lattice~\cite{Goddard:1986bp}. Due to this analogy, we write elements of the basic representation in a Fock space notation built on top of a ground state $|0\rangle$ (that is invariant under $T^A_0$ and $\{ L_{-1}, L_0, L_1\}$ as well as annihilated by $T_n^A$ for $n>0$) by acting with the negative mode generators 
\begin{align}
\cdots  T_{-n_2}^{A_2} T_{-n_1}^{A_1} | 0 \rangle\,,
\end{align}
with $n_i>1$. There is an intricate structure of null states in this Fock space whose removal yields an irreducible representation of E$_9$ on which also the $L_n$ act. Derivatives $\partial_M$ with respect to $Y^M$ are valued in the dual representation to the coordinates and written as bra vectors $\langle \partial |=\langle e^M| \partial_M $, where $\langle e^M|$ is a basis of the dual basic representation.  
The coordinate dependence of all fields and gauge parameters, denoted here collectively by $\phi_i$, is restricted by the section constraint \cite{Bossard:2017aae}
\begin{equation}\label{eq:SC}
\eta_{0\kern0.66pt\alpha\kern-0.2pt\beta}\langle\partial\phi_1|T^\alpha\kern-0.88pt\otimes\kern-0.8pt\langle\partial\phi_2|T^\beta
\hspace{-1.0mm}  =\hspace{-0.8mm} \langle\partial\phi_{2}|\kern-0.8pt\otimes\kern-0.8pt\langle\partial\phi_{1}| {-}\langle\partial\phi_{1}|\kern-0.8pt\otimes\kern-0.8pt\langle\partial\phi_{2}|\kern-1pt \; .
\end{equation}
An analogous condition also applies to second derivatives of a single field $\phi$.
The section constraint implies 
that all fields and parameters only depend on a finite subset of the $Y^M$.
Choosing any such subset breaks the manifest E$_9$-invariance. Besides the dependence on the `internal' coordinates $Y^M$ all fields also depend on the two `external' coordinates $x^\mu$ with $\mu=0,1$.
E$_9$ exceptional field theory becomes equivalent to either eleven-dimensional or type IIB supergravity upon choosing one of the (maximal) solutions to \eqref{eq:SC}.
In this letter, we focus on the internal sector of the theory that only involves derivatives with respect to the internal coordinates~$Y^M$.

Gauge symmetries act on fields by the so-called generalized Lie derivative.
It is defined by its action on a `generalized vector' $|V\rangle$ in the basic module
\begin{align}\label{eq:genLie}
\mathcal{L}_{|\Lambda\rangle,\,\Sigma} \, |V\rangle =\ & 
\Lambda^M
\partial_M |V \rangle - \eta_{0\, \alpha\beta} \langle \partial  | T^\alpha | \Lambda \rangle \, T^\beta | V\rangle \\\nonumber&
- \langle \partial  |\Lambda\rangle | V\rangle
- \eta_{-1\, \alpha\beta} {\rm Tr}(\Sigma\, T^\alpha) \, T^\beta | V\rangle
\,,
\end{align}
where the gauge parameter $|\Lambda\rangle$ is also a generalized vector, $\Lambda^M = \langle e^M| \Lambda\rangle$ and the derivatives in the second and third term act on 
$|\Lambda\rangle$. The parameter
$\Sigma$ is a so-called ancillary gauge parameter which is required 
for closure of the gauge algebra \cite{Hohm:2013jma,Bossard:2017aae}.
It can be written as a sum of tensor products of ket and bra vectors, with the bra vectors algebraically constrained as in \eqref{eq:SC}.

There are two types of scalar fields in E$_9$ exceptional field theory~\cite{Bossard:2021jix}.
The first  type corresponds to the infinitely many dualisations of the 128 propagating degrees of freedom in $D=2$ maximal supergravity~\cite{Nicolai:1987kz} and they are associated with the quotient of the Kac--Moody group E$_9$ by its maximal `compact' 
subgroup K(E$_9$). We represent them by a hermitian generalized metric $\mathcal{M}$ and a special role is played by the field $\rho$ that is the component along the Virasoro generator $L_0$. 
The second type is given by a so-called constrained scalar field $\langle \chi|$, where `constrained' refers to the fact that it can replace $\langle \partial \phi_i|$ in the section constraint \eqref{eq:SC} and therefore there are effectively at most nine independent components of $\langle \chi|$ that are non-vanishing.

Out of the scalar fields one can construct an $\mathfrak{e}_9$-valued current $\langle \mathcal{J}_\alpha|$ via the usual Maurer--Cartan derivative $\mathcal{M}^{-1} \partial_M \mathcal{M}$, as well as a shifted current $\langle \mathcal{J}^{-}_\alpha|$ in which the mode numbers are shifted by one negative unit and whose $\dK$-component is the constrained scalar $\langle \chi|$. 
The transformation of $\langle \chi|$ under rigid E$_9$ involves the components of $\langle \mathcal{J}_\alpha|$ such that $\langle \mathcal{J}^{-}_\alpha|$ transforms as a tensor.

The E$_9$ exceptional field theory potential is  bilinear in these two currents~\cite{Bossard:2018utw}
\begin{align}\label{eq:exftpot}
&\rho\kern.5ptV_{\kern-1pt\text{\tiny ExFT}} =
\tfrac14\eta_0^{\alpha\beta}\kern-.5pt\langle\mathcal{J}_\alpha|\kern-.5pt\mathcal{M}^{-1}\kern-1pt|\mathcal{J}_\beta\rangle
\kern-1pt-\kern-1pt\rho^{-1}\!\langle\partial\kern-.5pt\rho|T^\alpha\kern-1.5pt\mathcal{M}^{-1}\kern-1pt|\mathcal{J}_\alpha\rangle
\\[.5ex]&\nonumber
-\tfrac12\langle\mathcal{J}_\alpha|T^\beta\kern-1.5pt\mathcal{M}^{-1}\kern-.5ptT^\alpha{}^\dagger|\mathcal{J}_\beta\rangle 
+\tfrac12\rho^2\langle\mathcal{J}^-_\alpha|T^\beta\kern-1.5pt\mathcal{M}^{-1}T^\alpha{}^\dagger|\mathcal{J}^-_\beta\rangle
\, , 
\end{align}
and is invariant under gauge transformations up to a total derivative. We stress that the standard factor $\sqrt{-g}$ of the $D=2$ integration measure is absorbed into $V_{\text{\tiny ExFT}}$.
We also use the notation  $|\mathcal{J}_\alpha\rangle = (\langle\mathcal{J}_\alpha|)^\dagger$ and similarly for other bra vectors in the following.

\subsection{Generalized Scherk--Schwarz ansatz}

Generalized Scherk--Schwarz reductions \cite{Lee:2014mla,Cassani:2019vcl,Hohm:2014qga,Inverso:2017lrz} give a factorization ansatz of the $Y^M$ dependence of all fields (subject to the section constraint), such that the dynamics reduce to those of a gauged (maximal) supergravity and all solutions of the latter uplift to solutions of the full theory.
They are mainly encoded in a twist matrix $\,\mathcal{U}(Y)\in\mathrm{E}_9$, where $\mathcal{U}(Y)$ decomposes into $\twistrho(Y)^{-L_0}$ and an element of the loop group over E$_8$.
In particular, we define the $\mathfrak{e}_9$-valued  Weitzenb\"ock connection 
\begin{equation}  \langle W_\alpha| \otimes T^\alpha =\twistrho^{-1}  \langle e^M | \,\mathcal{U}^{-1} \otimes \partial_M \mathcal{U} \, \mathcal{U}^{-1} \; . \end{equation}
The tensor product $\otimes$ indicates that the bra vectors are not acted upon by the operators on its right.

The gauge transformations \eqref{eq:genLie} with the reduction ansatz \cite{Bossard:2017aae}
\begin{equation}\label{eq:gSSgauge}
|\Lambda\rangle = \twistrho^{-1} \mathcal{U}^{-1} |\lambda\rangle\,,\quad
\Sigma = \twistrho\, \mathcal{U}^{-1}T^\alpha|\lambda\rangle\langle W^+_\alpha|\,\mathcal{U} \,,
\end{equation}
must reduce to those of a gauged supergravity. Here,  $|\lambda\rangle$ is only allowed to depend on the external coordinates $x^\mu$.
This requirement translates to a differential constraint on the twist matrix.
In analogy with $\langle\mathcal{J}_\alpha^{-}|$, we define $\langle W^\pm_\alpha|$ as the Weitzenb\"ock connection with mode number shifted by $\pm1$ and whose central $\dK$-components $\langle w^\pm|$ are independent functions of $Y^M$, constrained in the same way as $\langle\chi|$. While $\langle w^\pm|$ were not considered in~\cite{Bossard:2017aae}, they are necessary to describe the most general Scherk--Schwarz ansatz and ensure manifest rigid E$_9$ covariance.
One then computes
\begin{align}
\twistrho\,\mathcal{U}\,\left(\,\mathcal{L}_{|\Lambda\rangle,\,\Sigma} \, |V\rangle\,\right)=\ & 
\eta_{-1\,\alpha\beta}\langle\theta |T^\alpha|\lambda\rangle 
\, T^\beta |v\rangle
\\\nonumber&
+\eta_{ 0\,\alpha\beta}\langle\vartheta|T^\alpha|\lambda\rangle 
\, T^\beta |v\rangle\,,
\end{align}
with $|V\rangle = \twistrho^{-1} \mathcal{U}^{-1} |v\rangle$, where $|v\rangle$ is $Y^M$-independent and
\begin{align}
\label{eq:embs}
\langle \theta | = - \langle W^+_\alpha | T^\alpha \,,\quad \langle \vartheta | =  \langle W_\alpha |T^\alpha\,.
\end{align}
Consistency of the truncation requires $\langle\theta|$ and $\langle\vartheta|$ to be constant, in which case they are identified with the components of the embedding tensor of two-dimensional gauged maximal supergravity \cite{Samtleben:2007an,Bossard:2017aae}.
The closure of the gauge algebra in supergravity is ensured by the so-called quadratic constraint~ \cite{Nicolai:2000sc,deWit:2002vt}. In the generalized Scherk--Schwarz ansatz this follows from closure of the exceptional field theory gauge algebra for both parameters $|\Lambda\rangle$ and $\Sigma$ in \eqref{eq:gSSgauge}. We have checked that the additional necessary condition on $\Sigma$ is automatically satisfied.

The reduction ansatz for standard scalar fields follows the ones of lower-rank exceptional field theories:
\begin{subequations}
\label{eq:gSSall}
\begin{align}
\label{eq:gSS}
\mathcal{M}(x,Y) &= \mathcal{U}^\dagger(Y) M(x) \mathcal{U}(Y)\,, \\
\rho(x,Y) &= \twistrho(Y) \sugrarho(x)\,,
\end{align}
\end{subequations}
where $M(x)$ and $\sugrarho(x)$ encode the scalar fields of (gauged) maximal supergravity.
Equation \eqref{eq:gSSall} needs to be supplemented by a reduction ansatz for the constrained scalar $\langle \chi |$. 
This is determined such that the shifted current splits into
\begin{align}\label{eq:gSSchi}
\langle \mathcal{J}^-_\alpha | \, \mathcal{U}^{-1} &\otimes \,\mathcal{U} T^\alpha \mathcal{U}^{-1} =
\\\nonumber
&\langle W^-_\alpha | \otimes T^\alpha   + \sugrarho^{-2}   \langle W^+_\alpha | \otimes M^{-1}  T^{\alpha \dagger} M 
\,.
\end{align}

A non-vanishing $\langle\vartheta|$ induces a gauging of $L_0$, which is only an on-shell symmetry and the resulting gauged supergravities do not admit a Lagrangian description, in analogy with trombone gaugings in higher dimensions~\cite{LeDiffon:2008sh}.
We will henceforth focus on Lagrangian gaugings, so that $\langle\vartheta|=0$.
One can then choose $\langle W^-_\alpha | T^\alpha  = 0 $ without loss of generality, thereby simplifying the final expression of the scalar potential.
By plugging the ansatz~\eqref{eq:gSSall} and \eqref{eq:gSSchi} into the potential of exceptional field theory \eqref{eq:exftpot}, we   compute the scalar potential of two-dimensional gauged maximal supergravity:
\begin{align}
\label{eq:Vpot}
\sugrarho \Vpot &=  \frac1{2\sugrarho^2} \langle \theta | M^{-1} | \theta \rangle  + \frac12 \eta_{-2\, \alpha\beta} \langle \theta | T^\alpha  M^{-1} T^{\beta \dagger} | \theta\rangle \nn\\
& \quad+ \frac{\sugrarho^2}2   \eta_{-4\, \alpha\beta} \langle \theta | T^\alpha M^{-1} T^{\beta \dagger} | \theta \rangle\,,
\end{align}
up to total derivatives. It is non-trivial that the potential can be fully expressed in terms of the quantities~\eqref{eq:embs}. The expression~\eqref{eq:Vpot} is one of the two main results reported in this letter. 
It defines, for the first time, the scalar potential of all two-dimensional gauged maximal supergravities admitting a geometric uplift to higher dimensions. As a cross-check, we have verified that it reproduces the potential of all three-dimensional gauged maximal supergravities also admitting an uplift.

\subsection{\texorpdfstring{Consistent Kaluza--Klein truncation on \boldmath$S^8$}{Consistent Kaluza--Klein truncation on S8}}

In order to illustrate the usefulness of the generalized Scherk--Schwarz procedure and of the general scalar potential~\eqref{eq:Vpot}, we now construct the consistent truncation of type IIA supergravity on $S^8$ (or equivalently, of eleven-dimensional supergravity on $S^8\times S^1$). It leads to a gauging of $D=2$ maximal supergravity that includes an SO(9) subgroup of E$_9$ that is not contained in E$_8$. This case relates to previous studies~\cite{Nicolai:2000zt,Ortiz:2012ib} using a different approach. To define the gauging we must give the twist matrix $\mathcal{U}$ in~\eqref{eq:gSS} whose Weitzenb\"ock connection determines the embedding tensor components~\eqref{eq:embs}. The corresponding ansatz for the twist matrix involves an SL(9) subgroup of E$_{9}$  containing the SO(9) gauge group.
This SL(9) is conjugate under E$_9$ to the one that acts on the $T^9$ compactification of $D=11$ supergravity.
The two share a common GL(8) subgroup containing  the  structure group of the $S^8$ compactification manifold. 

The dual of the basic representation decomposes as
\begin{align}
\label{eq:branch9}
 \overline{\bf 9}_{\frac{4}{9}} &\oplus {\bf 36}_{\frac{7}{9}} \oplus \overline{\bf 126}_{\frac{10}9} \oplus \left(\overline{\bf 9}\oplus {\bf 315}\right)_{\frac{13}{9}} \nn\\
&\oplus \left( {\bf 36} \oplus {\bf 45} \oplus {\bf 720}\right)_\frac{16}{9} \oplus \ldots
\end{align}
under this SL(9), where the subscripts denote the eigenvalues with respect to a redefined 
Virasoro generator $\LL_0$. It is determined such that it commutes with SL(9) instead of $E_8$.  
We write the basis vectors of the first two SL(9) representations in \eqref{eq:branch9} as \footnote{Normalizations in equations \eqref{eq:lowest bras}, \eqref{ansatz for r and s} and \eqref{eq:thetas} have been changed with respect to the published version of the paper.}
\begin{align}\label{eq:lowest bras}
\langle 0 |_I \,,\quad \langle \tfrac13|^{IJ} = 
\tfrac{1}{7}\langle 0|_K\T_{1/3}^{IJK} \,,
\end{align}
where $I,J$ are fundamental SL(9) indices and $\T_{1/3}^{IJK}$  ($\tfrac13$ being the $\LL_0$-eigenvalue) is the first raising operator in E$_9$ decomposed under this SL(9) and is fully antisymmetric in its indices.
The solution to the section constraint that is relevant for our examples consists in breaking SL(9) $\to$ SL(8) and keeping eight out of the ${\bf 36}_{7/9} \to {\bf 8} \oplus{\bf 28} $ components of $\langle \frac13|^{IJ}$. For a further embedding in $D=11$ one can add one more Kaluza--Klein circle whose coordinate is the singlet in ${\bf 45}_{16/9}\to {\bf 1}\oplus {\bf 8}\oplus {\bf 36}$.

The E$_9$ representation~\eqref{eq:branch9} not only governs the coordinates but also the embedding tensor components. We find a generalized Scherk--Schwarz ansatz for an embedding tensor defined as a symmetric tensor $\Theta_{IJ}$ in the ${\bf 45}_{16/9}$. The gauge group stabilizes the embedding tensor and when $\Theta_{IJ}= \coup \delta_{IJ} $, with $\coup$ the gauge coupling, we get SO(9) $\subset$ SL(9) gauged supergravity.

We choose the (inverse of the) twist matrix as
\begin{align}
\label{eq:twist}
\mathcal{U}^{-1} =   \twistrho^{\LL_0}  e^{\twistsigma \dK} \,\mathsf{u}^{-1}\,,
\end{align}
where $\mathsf{u}$ is an element of SL(9).
Computing the Weitzenb\"ock connection corresponding to the twist matrix~\eqref{eq:twist}, one can work out the components of~\eqref{eq:embs}. They simplify to finite expressions which are still a bit unwieldy but simplify further when using the standard sphere reduction ansatz~\cite{Lee:2014mla,Hohm:2014qga} for the SL(9) matrix $\mathsf{u}$. The latter can be written using nine embedding coordinates $y^I$ of a round $S^8$ in a nine-dimensional ambient space as
\begingroup
\allowdisplaybreaks
\begin{subequations}
\begin{align}
(\mathsf{u}^{-1})^{i}{}_I &= (\det g)^{1/9} \left( g^{ij} \partial_j y_I + c^i y_I\right)\,,\\
(\mathsf{u}^{-1})^{0}{}_I &= (\det g)^{-7/{18}} y_I\, . 
\end{align}
\end{subequations}
\endgroup
Here $g_{ij}$ is the induced metric on $S^8$ and we have split $I=(0,i)$; $c^i$ is the 7-form type IIA gauge potential, satisfying $\partial_i \left((\det g)^{1/2}c^i\right) =7 (\det g)^{1/2}$. 
The solutions for $\twistrho$ and $\twistsigma$ appearing in the twist matrix are given by \begin{equation}\label{ansatz for r and s}
\twistrho = (\det g)^{1/2}
\;  , \quad
 e^\twistsigma = \coup (\det g)^{7/18}\,.
\end{equation}
This ansatz requires $\langle w^+|=0$. Alternatively, one could reabsorb $c^i$ into a non-vanishing $\langle w^+|$, something that is not possible for lower-dimensional spheres and is related to the fact that the eleven-dimensional uplift of $c^i$ is a component of the dual graviton.

With these choices we obtain the following embedding tensors
\begin{align}
\label{eq:thetas}
\langle \theta | = - \frac{\coup}{8\textbf{}} \, \delta_{IJ} \langle \tfrac13 |^{KI} \T^J_{1\, K}\ \,, \quad
\langle \vartheta | =0 
\,,
\end{align}
which reproduce the embedding tensor of the SO(9) gauging.
These expressions straightforwardly generalize to $\mathrm{SO}(p,q)$ and $\mathrm{CSO}(p,q,r)$ gaugings, corresponding to other signatures of $\Theta_{IJ}$ in the ${\bf 45}_{16/9}$~\cite{Hull:1984qz,Samtleben:2007an,Ortiz:2012ib}.

For evaluating the potential~\eqref{eq:Vpot} on~\eqref{eq:thetas} we must also parametrize the supergravity scalar fields, i.e. $M(x)=V^\dagger V$ in~\eqref{eq:gSS}, where $V$ is the coset representative on E$_9$/K(E$_9)$. This takes a form similar to~\eqref{eq:twist} 
\begin{equation}
\label{eq:Vminus1}
V^{-1} = 
\cdots\, e^{h^J{}_I \T^I_{-1\, J}} 
e^{\frac16 a^{IJK} \T_{-\frac13\, IJK}} \sugrarho^{\LL_0} e^{\sugrasigma \dK} \,\V^{-1} \,,
\end{equation}
where now $\V$ labels the supergravity fields in our SL(9) and $a^{IJK}$ is anti-symmetric in its indices and couples to the first lowering generator of E$_9$ outside the loop algebra of SL(9).
The fields associated to all generators not shown explicitly in~\eqref{eq:Vminus1} drop out of the potential, including the one associated to $\T^{IJK}_{-2/3 }$.
Substituting this into the general potential~\eqref{eq:Vpot} leads to the following scalar potential for SO(9) gauged supergravity 
\begingroup
\allowdisplaybreaks
\begin{widetext}
\begin{align}
\label{eq:pot9}
  \Vpot &= \frac{\coup^2 e^{2\sugrasigma}}{2 } \sugrarho^{5/9} \delta_{IJ} \delta_{KL}  \Biggl( \left( 2\M^{IK} \M^{JL} -\M^{IJ} \M^{KL} \right)+\frac12  \sugrarho^{-2/3}   \Big( a^{IPQ} a^{KRS} \M^{JL} \M_{PR} \M_{QS} - 2 a^{IKP} a^{JLQ} \M_{PQ} \Big)\nn\\
& \quad + 2 \sugrarho^{-2} h^I{}_P h^K{}_Q \M^{Q[P} \M^{J]L}
+  \sugrarho^{-8/3}  a^{IPR} h^J{}_P a^{KQS} h^L{}_Q \M_{RS} \nn\\
&\quad + \frac{\sugrarho^{-2} }{72}  h^J{}_P a^{KQ_1Q_2} a^{LQ_3Q_4} a^{Q_5Q_6Q_7} \varepsilon_{Q_1\ldots Q_9} \M^{IQ_8} \M^{PQ_9}\nn\\
&\quad + \frac38  \sugrarho^{-4/3}  a^{I[M_1M_2} a^{M_3M_4]J} a^{K[N_1N_2} a^{N_3N_4]L}  \M_{M_1N_1} \M_{M_2N_2} \M_{M_3N_3} \M_{M_4N_4} \nn\\
&\quad +  \frac{ \sugrarho^{-2}}{2\cdot 144^2}  a^{IN_1N_2} a^{JN_3N_4} a^{N_5N_6N_7} \varepsilon_{N_1\ldots N_9}   a^{KP_1P_2} a^{LP_3P_4} a^{P_5P_6P_7} \varepsilon_{P_1\ldots P_9} \M^{N_8P_8} \M^{N_9P_9} \nn\\
&\quad + \frac{ \sugrarho^{-8/3}}{576}  a^{IRP} h^J{}_R a^{KN_1N_2}a^{LN_3N_4} a^{N_5N_6N_7}a^{N_8N_9Q} \varepsilon_{N_1\ldots N_9}  \M_{PQ}\nn\\
&\quad + \frac{\sugrarho^{-8/3}}{1152^2}  a^{IN_1N_2}a^{JN_3N_4} a^{N_5N_6N_7}a^{N_8N_9Q} \varepsilon_{N_1\ldots N_9} a^{KP_1P_2}a^{LP_3P_4} a^{P_5P_6P_7}a^{P_8P_9S} \varepsilon_{P_1\ldots P_9}  \M_{QS} \Biggr) \,.
\end{align}
\end{widetext}
\endgroup
This potential agrees with the one that can be deduced from~\cite[Eq.~(4.22)]{Ortiz:2012ib} up to conventions. The main result here is the constructive proof that appropriate extremization of this potential yields solutions that all uplift to vacua of eleven-dimensional supergravity. Here, $\M=\V^\dagger \V$ encodes the metric on $S^8$, the dilaton and the type-IIA seven-form, while $a^{IJK}$ encodes the type-IIA two-form and five-form. 
One can straightforwardly uplift further to $D=11$, but the reader should be warned that $a^{IJK}$ is \textit{not} the three-form on the nine-dimensional space.
The fields $h^I{}_J$ are auxiliary fields and they only appear through the anti-symmetric combination $\delta_{P[I} h^P{}_{J]}$. Integrating them out generates the two-dimensional Yang--Mills term for SO(9). The 128 propagating degrees of freedom are described by $\M^{IJ}$ and $a^{IJK}$~\cite{Ortiz:2012ib}.

When looking for ``vacuum'' solutions based on the scalar potential \eqref{eq:pot9}, one must take into account that the conformal factor $\sugrasigma$ of the $D=2$ metric as well as the dilaton $\sugrarho$ are typically running.
One must then extremize only with respect to the loop scalars to find dilaton-supported configurations.
The simplest extremum is given by $a^{IJK}=h^I{}_J=0$ and $\mathsf{m}_{IJ}=\delta_{IJ}$.
It consistently uplifts to the warped AdS$_2\times S^8\times S^1$ half-BPS solution in $D=11$ ~\cite{Nicolai:2000zt,Ortiz:2012ib}.
Supersymmetry implies stability of this solution, and one indeed checks that despite some negative signs in the potential the appropriate Breitenlohner--Freedman bound is respected for all modes in the two-dimensional theory.

The results of this letter open a new window on the study of AdS$_2$ vacua and matrix model holography.
The generalized Scherk--Schwarz ansatz (\ref{eq:gSSgauge}), (\ref{eq:gSSall}) and the scalar potential \eqref{eq:Vpot} allow for a systematic search of new consistent truncations to two-dimensional gauged supergravities with interesting extrema. 
Besides the analysis of the two-dimensional fluctuations, the explicit uplift ansatz enables one to analyse the full Kaluza--Klein spectrum in eleven dimensions using the techniques developed in~\cite{Malek:2019eaz}.
This would provide a streamlined re-derivation of previous results~\cite{Sekino:1999av} and allow for a straightforward generalization to less symmetric vacua, thus paving the way to precision tests for AdS$_2$ holography.

\medskip
\begin{acknowledgments}
We wish to thank B. K\"onig, H. Nicolai and H. Samtleben for discussions. This work was  supported by the European Union’s Horizon 2020 research and innovation programme (grant agreement No 740209). AK and FC gratefully acknowledge the hospitality of Ecole Polytechnique while part of this work was carried out. 
\end{acknowledgments}


\begin{thebibliography}{46}%
\makeatletter
\providecommand \@ifxundefined [1]{%
 \@ifx{#1\undefined}
}%
\providecommand \@ifnum [1]{%
 \ifnum #1\expandafter \@firstoftwo
 \else \expandafter \@secondoftwo
 \fi
}%
\providecommand \@ifx [1]{%
 \ifx #1\expandafter \@firstoftwo
 \else \expandafter \@secondoftwo
 \fi
}%
\providecommand \natexlab [1]{#1}%
\providecommand \enquote  [1]{``#1''}%
\providecommand \bibnamefont  [1]{#1}%
\providecommand \bibfnamefont [1]{#1}%
\providecommand \citenamefont [1]{#1}%
\providecommand \href@noop [0]{\@secondoftwo}%
\providecommand \href [0]{\begingroup \@sanitize@url \@href}%
\providecommand \@href[1]{\@@startlink{#1}\@@href}%
\providecommand \@@href[1]{\endgroup#1\@@endlink}%
\providecommand \@sanitize@url [0]{\catcode `\\12\catcode `\$12\catcode
  `\&12\catcode `\#12\catcode `\^12\catcode `\_12\catcode `\%12\relax}%
\providecommand \@@startlink[1]{}%
\providecommand \@@endlink[0]{}%
\providecommand \url  [0]{\begingroup\@sanitize@url \@url }%
\providecommand \@url [1]{\endgroup\@href {#1}{\urlprefix }}%
\providecommand \urlprefix  [0]{URL }%
\providecommand \Eprint [0]{\href }%
\providecommand \doibase [0]{https://doi.org/}%
\providecommand \selectlanguage [0]{\@gobble}%
\providecommand \bibinfo  [0]{\@secondoftwo}%
\providecommand \bibfield  [0]{\@secondoftwo}%
\providecommand \translation [1]{[#1]}%
\providecommand \BibitemOpen [0]{}%
\providecommand \bibitemStop [0]{}%
\providecommand \bibitemNoStop [0]{.\EOS\space}%
\providecommand \EOS [0]{\spacefactor3000\relax}%
\providecommand \BibitemShut  [1]{\csname bibitem#1\endcsname}%
\let\auto@bib@innerbib\@empty
\bibitem [{\citenamefont {Maldacena}(1998)}]{Maldacena:1997re}%
  \BibitemOpen
  \bibfield  {author} {\bibinfo {author} {\bibfnamefont {J.~M.}\ \bibnamefont
  {Maldacena}},\ }\href {https://doi.org/10.1023/A:1026654312961} {\bibfield
  {journal} {\bibinfo  {journal} {Adv. Theor. Math. Phys.}\ }\textbf {\bibinfo
  {volume} {2}},\ \bibinfo {pages} {231} (\bibinfo {year} {1998})},\ \Eprint
  {https://arxiv.org/abs/hep-th/9711200} {arXiv:hep-th/9711200} \BibitemShut
  {NoStop}%
\bibitem [{\citenamefont {Gubser}\ \emph {et~al.}(1998)\citenamefont {Gubser},
  \citenamefont {Klebanov},\ and\ \citenamefont {Polyakov}}]{Gubser:1998bc}%
  \BibitemOpen
  \bibfield  {author} {\bibinfo {author} {\bibfnamefont {S.~S.}\ \bibnamefont
  {Gubser}}, \bibinfo {author} {\bibfnamefont {I.~R.}\ \bibnamefont
  {Klebanov}},\ and\ \bibinfo {author} {\bibfnamefont {A.~M.}\ \bibnamefont
  {Polyakov}},\ }\href {https://doi.org/10.1016/S0370-2693(98)00377-3}
  {\bibfield  {journal} {\bibinfo  {journal} {Phys. Lett. B}\ }\textbf
  {\bibinfo {volume} {428}},\ \bibinfo {pages} {105} (\bibinfo {year}
  {1998})},\ \Eprint {https://arxiv.org/abs/hep-th/9802109}
  {arXiv:hep-th/9802109} \BibitemShut {NoStop}%
\bibitem [{\citenamefont {Witten}(1998)}]{Witten:1998qj}%
  \BibitemOpen
  \bibfield  {author} {\bibinfo {author} {\bibfnamefont {E.}~\bibnamefont
  {Witten}},\ }\href {https://doi.org/10.4310/ATMP.1998.v2.n2.a2} {\bibfield
  {journal} {\bibinfo  {journal} {Adv. Theor. Math. Phys.}\ }\textbf {\bibinfo
  {volume} {2}},\ \bibinfo {pages} {253} (\bibinfo {year} {1998})},\ \Eprint
  {https://arxiv.org/abs/hep-th/9802150} {arXiv:hep-th/9802150} \BibitemShut
  {NoStop}%
\bibitem [{\citenamefont {Ooguri}\ and\ \citenamefont
  {Vafa}(2017)}]{Ooguri:2016pdq}%
  \BibitemOpen
  \bibfield  {author} {\bibinfo {author} {\bibfnamefont {H.}~\bibnamefont
  {Ooguri}}\ and\ \bibinfo {author} {\bibfnamefont {C.}~\bibnamefont {Vafa}},\
  }\href {https://doi.org/10.4310/ATMP.2017.v21.n7.a8} {\bibfield  {journal}
  {\bibinfo  {journal} {Adv. Theor. Math. Phys.}\ }\textbf {\bibinfo {volume}
  {21}},\ \bibinfo {pages} {1787} (\bibinfo {year} {2017})},\ \Eprint
  {https://arxiv.org/abs/1610.01533} {arXiv:1610.01533 [hep-th]} \BibitemShut
  {NoStop}%
\bibitem [{\citenamefont {Palti}(2019)}]{Palti:2019pca}%
  \BibitemOpen
  \bibfield  {author} {\bibinfo {author} {\bibfnamefont {E.}~\bibnamefont
  {Palti}},\ }\href {https://doi.org/10.1002/prop.201900037} {\bibfield
  {journal} {\bibinfo  {journal} {Fortsch. Phys.}\ }\textbf {\bibinfo {volume}
  {67}},\ \bibinfo {pages} {1900037} (\bibinfo {year} {2019})},\ \Eprint
  {https://arxiv.org/abs/1903.06239} {arXiv:1903.06239 [hep-th]} \BibitemShut
  {NoStop}%
\bibitem [{\citenamefont {Beisert}\ \emph {et~al.}(2012)\citenamefont {Beisert}
  \emph {et~al.}}]{Beisert:2010jr}%
  \BibitemOpen
  \bibfield  {author} {\bibinfo {author} {\bibfnamefont {N.}~\bibnamefont
  {Beisert}} \emph {et~al.},\ }\href
  {https://doi.org/10.1007/s11005-011-0529-2} {\bibfield  {journal} {\bibinfo
  {journal} {Lett. Math. Phys.}\ }\textbf {\bibinfo {volume} {99}},\ \bibinfo
  {pages} {3} (\bibinfo {year} {2012})},\ \Eprint
  {https://arxiv.org/abs/1012.3982} {arXiv:1012.3982 [hep-th]} \BibitemShut
  {NoStop}%
\bibitem [{\citenamefont {Zarembo}(2017)}]{Zarembo:2016bbk}%
  \BibitemOpen
  \bibfield  {author} {\bibinfo {author} {\bibfnamefont {K.}~\bibnamefont
  {Zarembo}},\ }\href {https://doi.org/10.1088/1751-8121/aa585b} {\bibfield
  {journal} {\bibinfo  {journal} {J. Phys. A}\ }\textbf {\bibinfo {volume}
  {50}},\ \bibinfo {pages} {443011} (\bibinfo {year} {2017})},\ \Eprint
  {https://arxiv.org/abs/1608.02963} {arXiv:1608.02963 [hep-th]} \BibitemShut
  {NoStop}%
\bibitem [{\citenamefont {Samtleben}(2008)}]{Samtleben:2008pe}%
  \BibitemOpen
  \bibfield  {author} {\bibinfo {author} {\bibfnamefont {H.}~\bibnamefont
  {Samtleben}},\ }\href {https://doi.org/10.1088/0264-9381/25/21/214002}
  {\bibfield  {journal} {\bibinfo  {journal} {Class. Quant. Grav.}\ }\textbf
  {\bibinfo {volume} {25}},\ \bibinfo {pages} {214002} (\bibinfo {year}
  {2008})},\ \Eprint {https://arxiv.org/abs/0808.4076} {arXiv:0808.4076
  [hep-th]} \BibitemShut {NoStop}%
\bibitem [{\citenamefont {Trigiante}(2017)}]{Trigiante:2016mnt}%
  \BibitemOpen
  \bibfield  {author} {\bibinfo {author} {\bibfnamefont {M.}~\bibnamefont
  {Trigiante}},\ }\href {https://doi.org/10.1016/j.physrep.2017.03.001}
  {\bibfield  {journal} {\bibinfo  {journal} {Phys. Rept.}\ }\textbf {\bibinfo
  {volume} {680}},\ \bibinfo {pages} {1} (\bibinfo {year} {2017})},\ \Eprint
  {https://arxiv.org/abs/1609.09745} {arXiv:1609.09745 [hep-th]} \BibitemShut
  {NoStop}%
\bibitem [{\citenamefont {Lee}\ \emph {et~al.}(2017)\citenamefont {Lee},
  \citenamefont {Strickland-Constable},\ and\ \citenamefont
  {Waldram}}]{Lee:2014mla}%
  \BibitemOpen
  \bibfield  {author} {\bibinfo {author} {\bibfnamefont {K.}~\bibnamefont
  {Lee}}, \bibinfo {author} {\bibfnamefont {C.}~\bibnamefont
  {Strickland-Constable}},\ and\ \bibinfo {author} {\bibfnamefont
  {D.}~\bibnamefont {Waldram}},\ }\href
  {https://doi.org/10.1002/prop.201700048} {\bibfield  {journal} {\bibinfo
  {journal} {Fortsch. Phys.}\ }\textbf {\bibinfo {volume} {65}},\ \bibinfo
  {pages} {1700048} (\bibinfo {year} {2017})},\ \Eprint
  {https://arxiv.org/abs/1401.3360} {arXiv:1401.3360 [hep-th]} \BibitemShut
  {NoStop}%
\bibitem [{\citenamefont {Cassani}\ \emph {et~al.}(2019)\citenamefont
  {Cassani}, \citenamefont {Josse}, \citenamefont {Petrini},\ and\
  \citenamefont {Waldram}}]{Cassani:2019vcl}%
  \BibitemOpen
  \bibfield  {author} {\bibinfo {author} {\bibfnamefont {D.}~\bibnamefont
  {Cassani}}, \bibinfo {author} {\bibfnamefont {G.}~\bibnamefont {Josse}},
  \bibinfo {author} {\bibfnamefont {M.}~\bibnamefont {Petrini}},\ and\ \bibinfo
  {author} {\bibfnamefont {D.}~\bibnamefont {Waldram}},\ }\href
  {https://doi.org/10.1007/JHEP11(2019)017} {\bibfield  {journal} {\bibinfo
  {journal} {JHEP}\ }\textbf {\bibinfo {volume} {11}},\ \bibinfo {pages}
  {017}},\ \Eprint {https://arxiv.org/abs/1907.06730} {arXiv:1907.06730
  [hep-th]} \BibitemShut {NoStop}%
\bibitem [{\citenamefont {Hohm}\ and\ \citenamefont
  {Samtleben}(2015)}]{Hohm:2014qga}%
  \BibitemOpen
  \bibfield  {author} {\bibinfo {author} {\bibfnamefont {O.}~\bibnamefont
  {Hohm}}\ and\ \bibinfo {author} {\bibfnamefont {H.}~\bibnamefont
  {Samtleben}},\ }\href {https://doi.org/10.1007/JHEP01(2015)131} {\bibfield
  {journal} {\bibinfo  {journal} {JHEP}\ }\textbf {\bibinfo {volume} {01}},\
  \bibinfo {pages} {131}},\ \Eprint {https://arxiv.org/abs/1410.8145}
  {arXiv:1410.8145 [hep-th]} \BibitemShut {NoStop}%
\bibitem [{\citenamefont {Inverso}(2017)}]{Inverso:2017lrz}%
  \BibitemOpen
  \bibfield  {author} {\bibinfo {author} {\bibfnamefont {G.}~\bibnamefont
  {Inverso}},\ }\href {https://doi.org/10.1007/JHEP06(2021)148} {\bibfield
  {journal} {\bibinfo  {journal} {JHEP}\ }\textbf {\bibinfo {volume} {12}},\
  \bibinfo {pages} {124}},\ \bibinfo {note} {[Erratum: JHEP 06, 148 (2021)]},\
  \Eprint {https://arxiv.org/abs/1708.02589} {arXiv:1708.02589 [hep-th]}
  \BibitemShut {NoStop}%
\bibitem [{\citenamefont {Galli}\ \emph {et~al.}(2017)\citenamefont {Galli},\ and\ \citenamefont {Malek}}]{Galli:2022idq}%
  \BibitemOpen
  \bibfield  {author} {\bibinfo {author} {\bibfnamefont {M.}~\bibnamefont
  {Galli}}\ and\  \bibinfo {author} {\bibfnamefont {E.}~\bibnamefont
  {Malek}},\ }\href
  {https://doi.org/10.1007/JHEP09(2022)014} {\bibfield  {journal} {\bibinfo
  {journal} {JHEP}\ }\textbf {\bibinfo {volume} {09}},\ \bibinfo {pages}
  {014}},\ \Eprint {https://arxiv.org/abs/2206.03507} {arXiv:2206.03507
  [hep-th]} \BibitemShut {NoStop}%
\bibitem [{\citenamefont {Malek}\ and\ \citenamefont
  {Samtleben}(2020)}]{Malek:2019eaz}%
  \BibitemOpen
  \bibfield  {author} {\bibinfo {author} {\bibfnamefont {E.}~\bibnamefont
  {Malek}}\ and\ \bibinfo {author} {\bibfnamefont {H.}~\bibnamefont
  {Samtleben}},\ }\href {https://doi.org/10.1103/PhysRevLett.124.101601}
  {\bibfield  {journal} {\bibinfo  {journal} {Phys. Rev. Lett.}\ }\textbf
  {\bibinfo {volume} {124}},\ \bibinfo {pages} {101601} (\bibinfo {year}
  {2020})},\ \Eprint {https://arxiv.org/abs/1911.12640} {arXiv:1911.12640
  [hep-th]} \BibitemShut {NoStop}%
\bibitem [{\citenamefont {Guarino}\ \emph {et~al.}(2021)\citenamefont
  {Guarino}, \citenamefont {Malek},\ and\ \citenamefont
  {Samtleben}}]{Guarino:2020flh}%
  \BibitemOpen
  \bibfield  {author} {\bibinfo {author} {\bibfnamefont {A.}~\bibnamefont
  {Guarino}}, \bibinfo {author} {\bibfnamefont {E.}~\bibnamefont {Malek}},\
  and\ \bibinfo {author} {\bibfnamefont {H.}~\bibnamefont {Samtleben}},\ }\href
  {https://doi.org/10.1103/PhysRevLett.126.061601} {\bibfield  {journal}
  {\bibinfo  {journal} {Phys. Rev. Lett.}\ }\textbf {\bibinfo {volume} {126}},\
  \bibinfo {pages} {061601} (\bibinfo {year} {2021})},\ \Eprint
  {https://arxiv.org/abs/2011.06600} {arXiv:2011.06600 [hep-th]} \BibitemShut
  {NoStop}%
\bibitem [{\citenamefont {Nicolai}\ and\ \citenamefont
  {Samtleben}(2000)}]{Nicolai:2000zt}%
  \BibitemOpen
  \bibfield  {author} {\bibinfo {author} {\bibfnamefont {H.}~\bibnamefont
  {Nicolai}}\ and\ \bibinfo {author} {\bibfnamefont {H.}~\bibnamefont
  {Samtleben}},\ }\href {https://doi.org/10.22323/1.006.0014} {\bibfield
  {journal} {\bibinfo  {journal} {PoS}\ }\textbf {\bibinfo {volume}
  {tmr2000}},\ \bibinfo {pages} {014} (\bibinfo {year} {2000})}\BibitemShut
  {NoStop}%
\bibitem [{\citenamefont {Samtleben}\ and\ \citenamefont
  {Weidner}(2007)}]{Samtleben:2007an}%
  \BibitemOpen
  \bibfield  {author} {\bibinfo {author} {\bibfnamefont {H.}~\bibnamefont
  {Samtleben}}\ and\ \bibinfo {author} {\bibfnamefont {M.}~\bibnamefont
  {Weidner}},\ }\href {https://doi.org/10.1088/1126-6708/2007/08/076}
  {\bibfield  {journal} {\bibinfo  {journal} {JHEP}\ }\textbf {\bibinfo
  {volume} {08}},\ \bibinfo {pages} {076}},\ \Eprint
  {https://arxiv.org/abs/0705.2606} {arXiv:0705.2606 [hep-th]} \BibitemShut
  {NoStop}%
\bibitem [{\citenamefont {Ortiz}\ and\ \citenamefont
  {Samtleben}(2013)}]{Ortiz:2012ib}%
  \BibitemOpen
  \bibfield  {author} {\bibinfo {author} {\bibfnamefont {T.}~\bibnamefont
  {Ortiz}}\ and\ \bibinfo {author} {\bibfnamefont {H.}~\bibnamefont
  {Samtleben}},\ }\href {https://doi.org/10.1007/JHEP01(2013)183} {\bibfield
  {journal} {\bibinfo  {journal} {JHEP}\ }\textbf {\bibinfo {volume} {01}},\
  \bibinfo {pages} {183}},\ \Eprint {https://arxiv.org/abs/1210.4266}
  {arXiv:1210.4266 [hep-th]} \BibitemShut {NoStop}%
\bibitem [{\citenamefont {Almheiri}\ and\ \citenamefont
  {Polchinski}(2015)}]{Almheiri:2014cka}%
  \BibitemOpen
  \bibfield  {author} {\bibinfo {author} {\bibfnamefont {A.}~\bibnamefont
  {Almheiri}}\ and\ \bibinfo {author} {\bibfnamefont {J.}~\bibnamefont
  {Polchinski}},\ }\href {https://doi.org/10.1007/JHEP11(2015)014} {\bibfield
  {journal} {\bibinfo  {journal} {JHEP}\ }\textbf {\bibinfo {volume} {11}},\
  \bibinfo {pages} {014}},\ \Eprint {https://arxiv.org/abs/1402.6334}
  {arXiv:1402.6334 [hep-th]} \BibitemShut {NoStop}%
\bibitem [{\citenamefont {Maldacena}\ and\ \citenamefont
  {Stanford}(2016)}]{Maldacena:2016hyu}%
  \BibitemOpen
  \bibfield  {author} {\bibinfo {author} {\bibfnamefont {J.}~\bibnamefont
  {Maldacena}}\ and\ \bibinfo {author} {\bibfnamefont {D.}~\bibnamefont
  {Stanford}},\ }\href {https://doi.org/10.1103/PhysRevD.94.106002} {\bibfield
  {journal} {\bibinfo  {journal} {Phys. Rev. D}\ }\textbf {\bibinfo {volume}
  {94}},\ \bibinfo {pages} {106002} (\bibinfo {year} {2016})},\ \Eprint
  {https://arxiv.org/abs/1604.07818} {arXiv:1604.07818 [hep-th]} \BibitemShut
  {NoStop}%
\bibitem [{\citenamefont {Maldacena}\ \emph {et~al.}(2016)\citenamefont
  {Maldacena}, \citenamefont {Stanford},\ and\ \citenamefont
  {Yang}}]{Maldacena:2016upp}%
  \BibitemOpen
  \bibfield  {author} {\bibinfo {author} {\bibfnamefont {J.}~\bibnamefont
  {Maldacena}}, \bibinfo {author} {\bibfnamefont {D.}~\bibnamefont
  {Stanford}},\ and\ \bibinfo {author} {\bibfnamefont {Z.}~\bibnamefont
  {Yang}},\ }\href {https://doi.org/10.1093/ptep/ptw124} {\bibfield  {journal}
  {\bibinfo  {journal} {PTEP}\ }\textbf {\bibinfo {volume} {2016}},\ \bibinfo
  {pages} {12C104} (\bibinfo {year} {2016})},\ \Eprint
  {https://arxiv.org/abs/1606.01857} {arXiv:1606.01857 [hep-th]} \BibitemShut
  {NoStop}%
\bibitem [{\citenamefont {de~Wit}\ \emph {et~al.}(1988)\citenamefont {de~Wit},
  \citenamefont {Hoppe},\ and\ \citenamefont {Nicolai}}]{deWit:1988wri}%
  \BibitemOpen
  \bibfield  {author} {\bibinfo {author} {\bibfnamefont {B.}~\bibnamefont
  {de~Wit}}, \bibinfo {author} {\bibfnamefont {J.}~\bibnamefont {Hoppe}},\ and\
  \bibinfo {author} {\bibfnamefont {H.}~\bibnamefont {Nicolai}},\ }\href
  {https://doi.org/10.1016/0550-3213(88)90116-2} {\bibfield  {journal}
  {\bibinfo  {journal} {Nucl. Phys. B}\ }\textbf {\bibinfo {volume} {305}},\
  \bibinfo {pages} {545} (\bibinfo {year} {1988})}\BibitemShut {NoStop}%
\bibitem [{\citenamefont {Banks}\ \emph {et~al.}(1997)\citenamefont {Banks},
  \citenamefont {Fischler}, \citenamefont {Shenker},\ and\ \citenamefont
  {Susskind}}]{Banks:1996vh}%
  \BibitemOpen
  \bibfield  {author} {\bibinfo {author} {\bibfnamefont {T.}~\bibnamefont
  {Banks}}, \bibinfo {author} {\bibfnamefont {W.}~\bibnamefont {Fischler}},
  \bibinfo {author} {\bibfnamefont {S.~H.}\ \bibnamefont {Shenker}},\ and\
  \bibinfo {author} {\bibfnamefont {L.}~\bibnamefont {Susskind}},\ }\href
  {https://doi.org/10.1103/PhysRevD.55.5112} {\bibfield  {journal} {\bibinfo
  {journal} {Phys. Rev. D}\ }\textbf {\bibinfo {volume} {55}},\ \bibinfo
  {pages} {5112} (\bibinfo {year} {1997})},\ \Eprint
  {https://arxiv.org/abs/hep-th/9610043} {arXiv:hep-th/9610043} \BibitemShut
  {NoStop}%
\bibitem [{\citenamefont {Sekino}\ and\ \citenamefont
  {Yoneya}(2000)}]{Sekino:1999av}%
  \BibitemOpen
  \bibfield  {author} {\bibinfo {author} {\bibfnamefont {Y.}~\bibnamefont
  {Sekino}}\ and\ \bibinfo {author} {\bibfnamefont {T.}~\bibnamefont
  {Yoneya}},\ }\href {https://doi.org/10.1016/S0550-3213(99)00793-2} {\bibfield
   {journal} {\bibinfo  {journal} {Nucl. Phys. B}\ }\textbf {\bibinfo {volume}
  {570}},\ \bibinfo {pages} {174} (\bibinfo {year} {2000})},\ \Eprint
  {https://arxiv.org/abs/hep-th/9907029} {arXiv:hep-th/9907029} \BibitemShut
  {NoStop}%
\bibitem [{\citenamefont {Kanitscheider}\ \emph {et~al.}(2008)\citenamefont
  {Kanitscheider}, \citenamefont {Skenderis},\ and\ \citenamefont
  {Taylor}}]{Kanitscheider:2008kd}%
  \BibitemOpen
  \bibfield  {author} {\bibinfo {author} {\bibfnamefont {I.}~\bibnamefont
  {Kanitscheider}}, \bibinfo {author} {\bibfnamefont {K.}~\bibnamefont
  {Skenderis}},\ and\ \bibinfo {author} {\bibfnamefont {M.}~\bibnamefont
  {Taylor}},\ }\href {https://doi.org/10.1088/1126-6708/2008/09/094} {\bibfield
   {journal} {\bibinfo  {journal} {JHEP}\ }\textbf {\bibinfo {volume} {09}},\
  \bibinfo {pages} {094}},\ \Eprint {https://arxiv.org/abs/0807.3324}
  {arXiv:0807.3324 [hep-th]} \BibitemShut {NoStop}%
\bibitem [{\citenamefont {Ortiz}\ \emph {et~al.}(2014)\citenamefont {Ortiz},
  \citenamefont {Samtleben},\ and\ \citenamefont {Tsimpis}}]{Ortiz:2014aja}%
  \BibitemOpen
  \bibfield  {author} {\bibinfo {author} {\bibfnamefont {T.}~\bibnamefont
  {Ortiz}}, \bibinfo {author} {\bibfnamefont {H.}~\bibnamefont {Samtleben}},\
  and\ \bibinfo {author} {\bibfnamefont {D.}~\bibnamefont {Tsimpis}},\ }\href
  {https://doi.org/10.1007/JHEP12(2014)096} {\bibfield  {journal} {\bibinfo
  {journal} {JHEP}\ }\textbf {\bibinfo {volume} {12}},\ \bibinfo {pages}
  {096}},\ \Eprint {https://arxiv.org/abs/1410.0487} {arXiv:1410.0487 [hep-th]}
  \BibitemShut {NoStop}%
\bibitem [{\citenamefont {Anabal\'on}\ \emph {et~al.}(2013)\citenamefont
  {Anabal\'on}, \citenamefont {Ortiz},\ and\ \citenamefont
  {Samtleben}}]{Anabalon:2013zka}%
  \BibitemOpen
  \bibfield  {author} {\bibinfo {author} {\bibfnamefont {A.}~\bibnamefont
  {Anabal\'on}}, \bibinfo {author} {\bibfnamefont {T.}~\bibnamefont {Ortiz}},\
  and\ \bibinfo {author} {\bibfnamefont {H.}~\bibnamefont {Samtleben}},\ }\href
  {https://doi.org/10.1016/j.physletb.2013.10.049} {\bibfield  {journal}
  {\bibinfo  {journal} {Phys. Lett. B}\ }\textbf {\bibinfo {volume} {727}},\
  \bibinfo {pages} {516} (\bibinfo {year} {2013})},\ \Eprint
  {https://arxiv.org/abs/1310.1321} {arXiv:1310.1321 [hep-th]} \BibitemShut
  {NoStop}%
\bibitem [{\citenamefont {de~Wit}\ and\ \citenamefont
  {Nicolai}(1982)}]{deWit:1982bul}%
  \BibitemOpen
  \bibfield  {author} {\bibinfo {author} {\bibfnamefont {B.}~\bibnamefont
  {de~Wit}}\ and\ \bibinfo {author} {\bibfnamefont {H.}~\bibnamefont
  {Nicolai}},\ }\href {https://doi.org/10.1016/0550-3213(82)90120-1} {\bibfield
   {journal} {\bibinfo  {journal} {Nucl. Phys. B}\ }\textbf {\bibinfo {volume}
  {208}},\ \bibinfo {pages} {323} (\bibinfo {year} {1982})}\BibitemShut
  {NoStop}%
\bibitem [{\citenamefont {Hull}(1984{\natexlab{a}})}]{Hull:1984vg}%
  \BibitemOpen
  \bibfield  {author} {\bibinfo {author} {\bibfnamefont {C.~M.}\ \bibnamefont
  {Hull}},\ }\href {https://doi.org/10.1016/0370-2693(84)91131-6} {\bibfield
  {journal} {\bibinfo  {journal} {Phys. Lett. B}\ }\textbf {\bibinfo {volume}
  {142}},\ \bibinfo {pages} {39} (\bibinfo {year}
  {1984}{\natexlab{a}})}\BibitemShut {NoStop}%
\bibitem [{\citenamefont {G{\"{u}}naydin}\ \emph {et~al.}(1986)\citenamefont
  {G{\"{u}}naydin}, \citenamefont {Romans},\ and\ \citenamefont
  {Warner}}]{Gunaydin:1985cu}%
  \BibitemOpen
  \bibfield  {author} {\bibinfo {author} {\bibfnamefont {M.}~\bibnamefont
  {G{\"{u}}naydin}}, \bibinfo {author} {\bibfnamefont {L.~J.}\ \bibnamefont
  {Romans}},\ and\ \bibinfo {author} {\bibfnamefont {N.~P.}\ \bibnamefont
  {Warner}},\ }\href {https://doi.org/10.1016/0550-3213(86)90237-3} {\bibfield
  {journal} {\bibinfo  {journal} {Nucl. Phys. B}\ }\textbf {\bibinfo {volume}
  {272}},\ \bibinfo {pages} {598} (\bibinfo {year} {1986})}\BibitemShut
  {NoStop}%
\bibitem [{\citenamefont {Nicolai}\ and\ \citenamefont
  {Samtleben}(2001)}]{Nicolai:2000sc}%
  \BibitemOpen
  \bibfield  {author} {\bibinfo {author} {\bibfnamefont {H.}~\bibnamefont
  {Nicolai}}\ and\ \bibinfo {author} {\bibfnamefont {H.}~\bibnamefont
  {Samtleben}},\ }\href {https://doi.org/10.1103/PhysRevLett.86.1686}
  {\bibfield  {journal} {\bibinfo  {journal} {Phys. Rev. Lett.}\ }\textbf
  {\bibinfo {volume} {86}},\ \bibinfo {pages} {1686} (\bibinfo {year}
  {2001})},\ \Eprint {https://arxiv.org/abs/hep-th/0010076}
  {arXiv:hep-th/0010076} \BibitemShut {NoStop}%
\bibitem [{\citenamefont {de~Wit}\ \emph {et~al.}(2003)\citenamefont {de~Wit},
  \citenamefont {Samtleben},\ and\ \citenamefont {Trigiante}}]{deWit:2002vt}%
  \BibitemOpen
  \bibfield  {author} {\bibinfo {author} {\bibfnamefont {B.}~\bibnamefont
  {de~Wit}}, \bibinfo {author} {\bibfnamefont {H.}~\bibnamefont {Samtleben}},\
  and\ \bibinfo {author} {\bibfnamefont {M.}~\bibnamefont {Trigiante}},\ }\href
  {https://doi.org/10.1016/S0550-3213(03)00059-2} {\bibfield  {journal}
  {\bibinfo  {journal} {Nucl. Phys. B}\ }\textbf {\bibinfo {volume} {655}},\
  \bibinfo {pages} {93} (\bibinfo {year} {2003})},\ \Eprint
  {https://arxiv.org/abs/hep-th/0212239} {arXiv:hep-th/0212239} \BibitemShut
  {NoStop}%
\bibitem [{\citenamefont {Nicolai}\ and\ \citenamefont
  {Samtleben}(2005)}]{Nicolai:2004nv}%
  \BibitemOpen
  \bibfield  {author} {\bibinfo {author} {\bibfnamefont {H.}~\bibnamefont
  {Nicolai}}\ and\ \bibinfo {author} {\bibfnamefont {H.}~\bibnamefont
  {Samtleben}},\ }\href {https://doi.org/10.4310/PAMQ.2005.v1.n1.a8} {\bibfield
   {journal} {\bibinfo  {journal} {Q. J. Pure Appl. Math.}\ }\textbf {\bibinfo
  {volume} {1}},\ \bibinfo {pages} {180} (\bibinfo {year} {2005})},\ \Eprint
  {https://arxiv.org/abs/hep-th/0407055} {arXiv:hep-th/0407055} \BibitemShut
  {NoStop}%
\bibitem [{\citenamefont {Kleinschmidt}\ \emph {et~al.}(2022)\citenamefont
  {Kleinschmidt}, \citenamefont {K\"ohl}, \citenamefont {Lautenbacher},\ and\
  \citenamefont {Nicolai}}]{Kleinschmidt:2021agj}%
  \BibitemOpen
  \bibfield  {author} {\bibinfo {author} {\bibfnamefont {A.}~\bibnamefont
  {Kleinschmidt}}, \bibinfo {author} {\bibfnamefont {R.}~\bibnamefont
  {K\"ohl}}, \bibinfo {author} {\bibfnamefont {R.}~\bibnamefont
  {Lautenbacher}},\ and\ \bibinfo {author} {\bibfnamefont {H.}~\bibnamefont
  {Nicolai}},\ }\href {https://doi.org/10.1007/s00220-022-04342-9} {\bibfield
  {journal} {\bibinfo  {journal} {Commun. Math. Phys.}\ }\textbf {\bibinfo
  {volume} {392}},\ \bibinfo {pages} {89} (\bibinfo {year} {2022})},\ \Eprint
  {https://arxiv.org/abs/2102.00870} {arXiv:2102.00870 [math.RT]} \BibitemShut
  {NoStop}%
\bibitem [{\citenamefont {Bossard}\ \emph {et~al.}(2019)\citenamefont
  {Bossard}, \citenamefont {Ciceri}, \citenamefont {Inverso}, \citenamefont
  {Kleinschmidt},\ and\ \citenamefont {Samtleben}}]{Bossard:2018utw}%
  \BibitemOpen
  \bibfield  {author} {\bibinfo {author} {\bibfnamefont {G.}~\bibnamefont
  {Bossard}}, \bibinfo {author} {\bibfnamefont {F.}~\bibnamefont {Ciceri}},
  \bibinfo {author} {\bibfnamefont {G.}~\bibnamefont {Inverso}}, \bibinfo
  {author} {\bibfnamefont {A.}~\bibnamefont {Kleinschmidt}},\ and\ \bibinfo
  {author} {\bibfnamefont {H.}~\bibnamefont {Samtleben}},\ }\href
  {https://doi.org/10.1007/JHEP03(2019)089} {\bibfield  {journal} {\bibinfo
  {journal} {JHEP}\ }\textbf {\bibinfo {volume} {03}},\ \bibinfo {pages}
  {089}},\ \Eprint {https://arxiv.org/abs/1811.04088} {arXiv:1811.04088
  [hep-th]} \BibitemShut {NoStop}%
\bibitem [{\citenamefont {Bossard}\ \emph {et~al.}(2021)\citenamefont
  {Bossard}, \citenamefont {Ciceri}, \citenamefont {Inverso}, \citenamefont
  {Kleinschmidt},\ and\ \citenamefont {Samtleben}}]{Bossard:2021jix}%
  \BibitemOpen
  \bibfield  {author} {\bibinfo {author} {\bibfnamefont {G.}~\bibnamefont
  {Bossard}}, \bibinfo {author} {\bibfnamefont {F.}~\bibnamefont {Ciceri}},
  \bibinfo {author} {\bibfnamefont {G.}~\bibnamefont {Inverso}}, \bibinfo
  {author} {\bibfnamefont {A.}~\bibnamefont {Kleinschmidt}},\ and\ \bibinfo
  {author} {\bibfnamefont {H.}~\bibnamefont {Samtleben}},\ }\href
  {https://doi.org/10.1007/JHEP05(2021)107} {\bibfield  {journal} {\bibinfo
  {journal} {JHEP}\ }\textbf {\bibinfo {volume} {05}},\ \bibinfo {pages}
  {107}},\ \Eprint {https://arxiv.org/abs/2103.12118} {arXiv:2103.12118
  [hep-th]} \BibitemShut {NoStop}%
\bibitem [{\citenamefont {Julia}(1981)}]{Julia:1980gr}%
  \BibitemOpen
  \bibfield  {author} {\bibinfo {author} {\bibfnamefont {B.}~\bibnamefont
  {Julia}},\ }in\ \href@noop {} {\emph {\bibinfo {booktitle} {{Superspace and
  Supergravity}}}},\ \bibinfo {editor} {edited by\ \bibinfo {editor}
  {\bibfnamefont {S.~W.}\ \bibnamefont {Hawking}}\ and\ \bibinfo {editor}
  {\bibfnamefont {M.}~\bibnamefont {Rocek}}}\ (\bibinfo  {publisher} {Cambridge
  University Press},\ \bibinfo {year} {1981})\ pp.\ \bibinfo {pages}
  {331--350}\BibitemShut {NoStop}%
\bibitem [{\citenamefont {Cremmer}\ \emph {et~al.}(1998)\citenamefont
  {Cremmer}, \citenamefont {Julia}, \citenamefont {{L\"{u}}},\ and\
  \citenamefont {Pope}}]{Cremmer:1997ct}%
  \BibitemOpen
  \bibfield  {author} {\bibinfo {author} {\bibfnamefont {E.}~\bibnamefont
  {Cremmer}}, \bibinfo {author} {\bibfnamefont {B.}~\bibnamefont {Julia}},
  \bibinfo {author} {\bibfnamefont {H.}~\bibnamefont {{L\"{u}}}},\ and\
  \bibinfo {author} {\bibfnamefont {C.~N.}\ \bibnamefont {Pope}},\ }\href
  {https://doi.org/10.1016/S0550-3213(98)00136-9} {\bibfield  {journal}
  {\bibinfo  {journal} {Nucl. Phys. B}\ }\textbf {\bibinfo {volume} {523}},\
  \bibinfo {pages} {73} (\bibinfo {year} {1998})},\ \Eprint
  {https://arxiv.org/abs/hep-th/9710119} {arXiv:hep-th/9710119} \BibitemShut
  {NoStop}%
\bibitem [{\citenamefont {Julia}(1982)}]{Julia:1982gx}%
  \BibitemOpen
  \bibfield  {author} {\bibinfo {author} {\bibfnamefont {B.}~\bibnamefont
  {Julia}},\ }in\ \href@noop {} {\emph {\bibinfo {booktitle} {{American
  Mathematical Society summer seminar on Application of Group Theory in Physics
  and Mathematical Physics}}}}\ (\bibinfo {year} {1982}), preprint \href{https://lib-extopc.kek.jp/preprints/PDF/1982/8212/8212081.pdf}{LPTENS-82-22}\BibitemShut {NoStop}%
\bibitem [{\citenamefont {Nicolai}(1987)}]{Nicolai:1987kz}%
  \BibitemOpen
  \bibfield  {author} {\bibinfo {author} {\bibfnamefont {H.}~\bibnamefont
  {Nicolai}},\ }\href {https://doi.org/10.1016/0370-2693(87)91072-0} {\bibfield
   {journal} {\bibinfo  {journal} {Phys. Lett. B}\ }\textbf {\bibinfo {volume}
  {194}},\ \bibinfo {pages} {402} (\bibinfo {year} {1987})}\BibitemShut
  {NoStop}%
\bibitem [{\citenamefont {Bossard}\ \emph {et~al.}(2017)\citenamefont
  {Bossard}, \citenamefont {Cederwall}, \citenamefont {Kleinschmidt},
  \citenamefont {Palmkvist},\ and\ \citenamefont
  {Samtleben}}]{Bossard:2017aae}%
  \BibitemOpen
  \bibfield  {author} {\bibinfo {author} {\bibfnamefont {G.}~\bibnamefont
  {Bossard}}, \bibinfo {author} {\bibfnamefont {M.}~\bibnamefont {Cederwall}},
  \bibinfo {author} {\bibfnamefont {A.}~\bibnamefont {Kleinschmidt}}, \bibinfo
  {author} {\bibfnamefont {J.}~\bibnamefont {Palmkvist}},\ and\ \bibinfo
  {author} {\bibfnamefont {H.}~\bibnamefont {Samtleben}},\ }\href
  {https://doi.org/10.1103/PhysRevD.96.106022} {\bibfield  {journal} {\bibinfo
  {journal} {Phys. Rev. D}\ }\textbf {\bibinfo {volume} {96}},\ \bibinfo
  {pages} {106022} (\bibinfo {year} {2017})},\ \Eprint
  {https://arxiv.org/abs/1708.08936} {arXiv:1708.08936 [hep-th]} \BibitemShut
  {NoStop}%
\bibitem [{Note1()}]{Note1}%
  \BibitemOpen
  \bibinfo {note} {For $k=0$ the form is related to the standard invariant
  bilinear form when restricted to the actual Kac--Moody algebra $\protect
  \mathfrak {e}_9$.}\BibitemShut {Stop}%
\bibitem [{\citenamefont {Goddard}\ and\ \citenamefont
  {Olive}(1986)}]{Goddard:1986bp}%
  \BibitemOpen
  \bibfield  {author} {\bibinfo {author} {\bibfnamefont {P.}~\bibnamefont
  {Goddard}}\ and\ \bibinfo {author} {\bibfnamefont {D.~I.}\ \bibnamefont
  {Olive}},\ }\href {https://doi.org/10.1142/S0217751X86000149} {\bibfield
  {journal} {\bibinfo  {journal} {Int. J. Mod. Phys. A}\ }\textbf {\bibinfo
  {volume} {1}},\ \bibinfo {pages} {303} (\bibinfo {year} {1986})}\BibitemShut
  {NoStop}%
\bibitem [{\citenamefont {Hohm}\ and\ \citenamefont
  {Samtleben}(2013)}]{Hohm:2013jma}%
  \BibitemOpen
  \bibfield  {author} {\bibinfo {author} {\bibfnamefont {O.}~\bibnamefont
  {Hohm}}\ and\ \bibinfo {author} {\bibfnamefont {H.}~\bibnamefont
  {Samtleben}},\ }\href {https://doi.org/10.1007/JHEP09(2013)080} {\bibfield
  {journal} {\bibinfo  {journal} {JHEP}\ }\textbf {\bibinfo {volume} {09}},\
  \bibinfo {pages} {080}},\ \Eprint {https://arxiv.org/abs/1307.0509}
  {arXiv:1307.0509 [hep-th]} \BibitemShut {NoStop}%
\bibitem [{\citenamefont {Le~Diffon}\ and\ \citenamefont
  {Samtleben}(2009)}]{LeDiffon:2008sh}%
  \BibitemOpen
  \bibfield  {author} {\bibinfo {author} {\bibfnamefont {A.}~\bibnamefont
  {Le~Diffon}}\ and\ \bibinfo {author} {\bibfnamefont {H.}~\bibnamefont
  {Samtleben}},\ }\href {https://doi.org/10.1016/j.nuclphysb.2008.11.010}
  {\bibfield  {journal} {\bibinfo  {journal} {Nucl. Phys. B}\ }\textbf
  {\bibinfo {volume} {811}},\ \bibinfo {pages} {1} (\bibinfo {year} {2009})},\
  \Eprint {https://arxiv.org/abs/0809.5180} {arXiv:0809.5180 [hep-th]}
  \BibitemShut {NoStop}%
\bibitem [{Note2()}]{Note2}%
  \BibitemOpen
  \bibinfo {note} {Normalizations in equations \protect \eqref {eq:lowest
  bras}, \protect \eqref {ansatz for r and s} and \protect \eqref {eq:thetas}
  have been changed with respect to the published version of the
  paper.}\BibitemShut {Stop}%
\bibitem [{\citenamefont {Hull}(1984{\natexlab{b}})}]{Hull:1984qz}%
  \BibitemOpen
  \bibfield  {author} {\bibinfo {author} {\bibfnamefont {C.~M.}\ \bibnamefont
  {Hull}},\ }\href {https://doi.org/10.1016/0370-2693(84)90091-1} {\bibfield
  {journal} {\bibinfo  {journal} {Phys. Lett. B}\ }\textbf {\bibinfo {volume}
  {148}},\ \bibinfo {pages} {297} (\bibinfo {year}
  {1984}{\natexlab{b}})}\BibitemShut {NoStop}%
\end{thebibliography}


%

\end{document}